# High-frequency, scaled MoS$_2$ transistors

Daria Krasnozhon[1], Subhojit Dutta[2], Clemens Nyffeler[1], Yusuf Leblebici[1], Andras Kis[1]*

[1]Electrical Engineering Institute, École Polytechnique Fédérale de Lausanne (EPFL), CH-1015 Lausanne, Switzerland
[2]Department of Electrical Engineering, Indian Institute of Technology (IIT), Kharagpur, West Bengal 721302, India
*Correspondence should be addressed to: Andras Kis, andras.kis@epfl.ch

**Abstract**

The interest in MoS$_2$ for radio-frequency (RF) applications has recently increased[1,2]. However, little is known on the scaling behavior of transistors made from MoS$_2$ for RF applications, which is important for establishing performance limits for electronic circuits based on 2D semiconductors on flexible and rigid substrates. Here, we present a systematic study of top-gated trilayer MoS$_2$ RF transistors with gate lengths scaled down to 70 and 40 nm. In addition, by introducing "edge-contacted" injection of electrons[3] in trilayer MoS$_2$ devices, we decrease the contact resistance and as a result obtain the highest cutoff frequency of 6 GHz before the de-embedding procedure and 25 GHz after the de-embedding procedure.

**Introduction**

The family of transition metal dichalcogenides (TMDCs) has attracted a lot of attention in the last years due to their remarkable electronic, optical and mechanical properties[4,5,6,7]. Operation of MoS$_2$ FETs in the RF range has been demonstrated recently[1]. Here we present the scaled high-frequency trilayer MoS$_2$ FETs with improved characteristics. The presence of a band gap in MoS$_2$ makes it easier to reach current saturation. As a consequence, for both $f_T$ and $f_{max}$ we are able to see the typical 1/L behavior down to 40 nm. We also observe voltage gain for all our devices. In contrast to MoS$_2$, for graphene-based RF devices[8] $f_T$ exhibits a 1/L trend, while $f_{max}$ has a nonmonotonic behavior with the decreasing channel length, reaching a peak value around 200 nm. Such behavior can be ascribed to a competition between the gate resistance and the output conductance $g_d$ with decreasing gate length.

**Device fabrication**

MoS$_2$ FETs were fabricated from MoS$_2$ exfoliated onto a highly resistive intrinsic Si substrate covered with a 270 nm thick SiO$_2$ layer. Figure 1, (c) shows the schematic cross-sectional view of MoS$_2$ FET. Electrical contacts were patterned using electron-beam lithography and by depositing 70 nm thick gold electrodes. This was followed by an annealing step at 200 °C in order to remove resist residue and decrease the contact resistance. Atomic layer deposition (ALD) was used to deposit a 20 nm thick layer of HfO$_2$ as high-κ gate dielectric. Local top gates for controlling the current in the two branches were fabricated using another e-beam lithography step followed by evaporation of 10 nm/50 nm of Cr/Au. We fabricated devices with 70 nm and 40 nm gate lengths, and conventional source-drain contacts (Fig. 1).

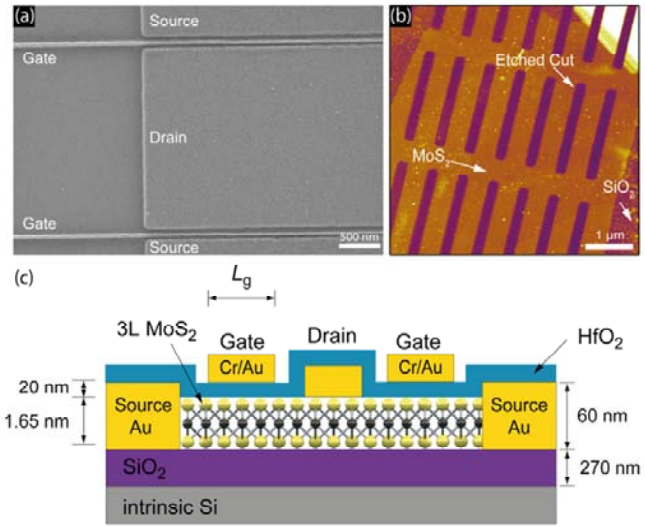

**Fig. 1** (a) Scanning electron microscopy image of top-gated MoS$_2$ field-effect transistors with 40 nm gate length, and conventional source-drain contacts, (b) Atomic force microscopy image of "edge-contacted" device to decrease contact resistance, showing typical cut pattern with cut widths of 200 nm and MoS$_2$ nanoribbon widths of 550 nm between each cut, (c) Schematic cross – sectional view of the RF – MoS$_2$ FET.

In order to enhance carrier injection, in some devices we also pattern MoS$_2$ in the contact area by introducing cuts, with the aim of increasing the total length of the MoS$_2$ edge under the metal electrode[3] (Fig. 1b). The device characteristics for 240 nm gate length used here for comparison purposes were originally reported in our previous work[1].

**Measurement setup**

Device characterization is performed at room temperature in an RF probe station (Cascade Microtech). DC and AC voltages are applied using bias tees connected to each probe head. DC currents are measured using an Agilent B2912A parameter analyzer. The AC excitation and subsequent S-parameter measurements are performed using an Agilent N5224A vector-network analyzer (VNA). High-frequency S-parameter measurements are carried out in the 0.1−50 GHz frequency range. The probe system is calibrated before the measurements of the device under test (DUT) in order to take into account any spurious contribution from connectors, cables, and the electrical environment of the DUT and to subtract it from the measured signal. This is done first by means of a calibration pad and in the second step using a set of dummy structures.

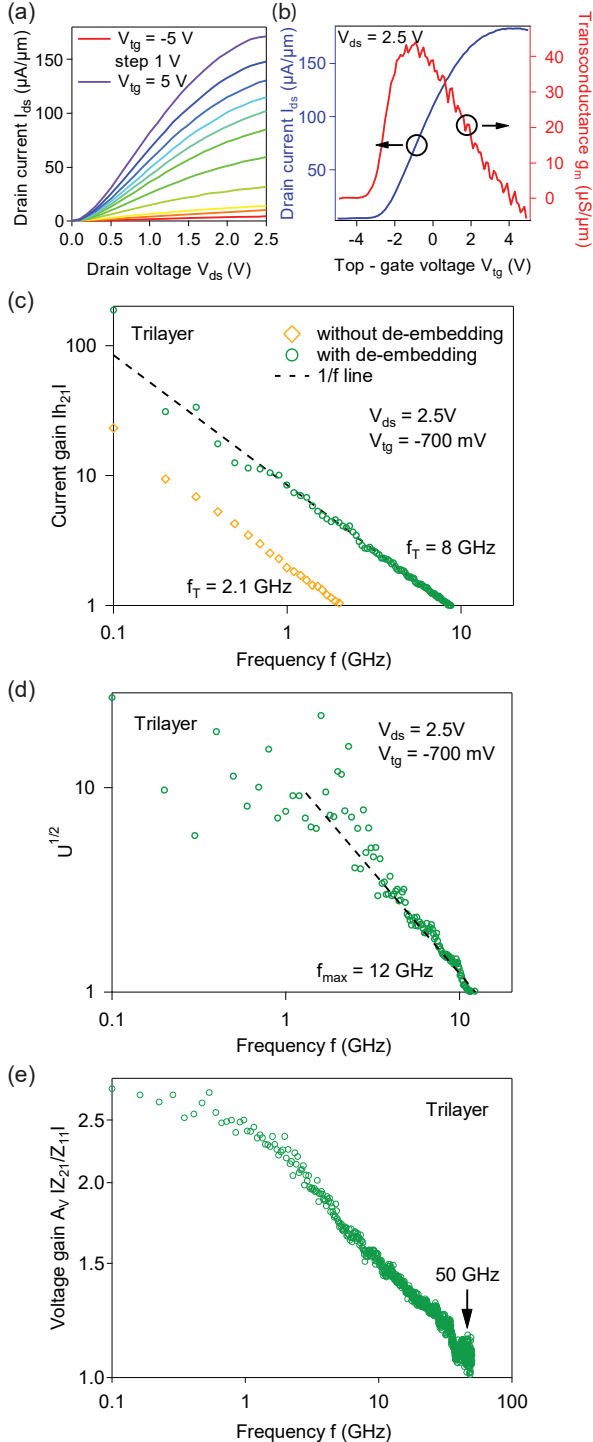

**Fig. 2** Device characteristics for 70 nm gate length and conventional source-drain contacts. (a) $I_{ds}$–$V_{ds}$ characteristics for the trilayer MoS$_2$ device, measured for different top – gate voltages. (b) Left: Transfer characteristics for the trilayer MoS$_2$ FET for $V_{ds}$ = 2.5 V. Right: Transconductance of trilayer MoS$_2$ derived from $I_{ds}$–$V_{tg}$ characteristics. (c) Small-signal current gain $h_{21}$ as a function of frequency for a device with 70 nm gate length based on trilayer MoS$_2$. The cut-off frequency $f_T$ = 8 GHz. (d) Mason's unilateral gain U as a function of frequency for device with a 70 nm gate length based on trilayer MoS$_2$. The maximum frequency of oscillations $f_{max}$ = 12 GHz. (e) Voltage gain $|Z_{21}/Z_{11}|$ as a function of frequency for the device with a 70 nm gate length based on trilayer MoS$_2$, showing gain up to 50 GHz.

The system was calibrated using the line – reflect – reflect – match method for the required frequency range and at low input power (typically −27 dBm) using a standard CSR-8 substrate. On-chip OPEN and SHORT structures were used to de-embed the parasitic effects, with the purpose of obtaining the intrinsic RF performance.

**Results**

In our previous studies we found that trilayer MoS$_2$ FETs have the best RF performance[1]. This motivated us to continue the research of scaled trilayer MoS$_2$ devices.

Transfer and output characteristics of 70 nm gate length trilayer MoS$_2$ are presented in Fig. 2 (a), (b). For this device, the on-current is ~180 μA/μm at $V_{ds}$ = 2.5 V with the intrinsic transconductance of 45 μS/μm. The specific contact geometry dictated by the need to interface RF transistors with the network analyzer and RF probes prohibits the fabrication of four or six contact devices, necessary to accurately measure the contact resistance and field-effect mobility in our devices. We can however, give an estimate of these quantities from the transfer characteristics, based on a simple model[1]. The dielectric constant of 20 nm HfO$_2$ deposited at 200 °C is 14, probably due to surface impurities trapped between the HfO$_2$ and MoS$_2$ layer. For 70 nm gate length trilayer MoS$_2$ we obtained μ = 8.3 cm$^2$/Vs and $R_c$ = 326 Ω or $R_c$ = 5.5 kΩ·μm with the channel width of 17 μm. We see drain current saturation in this short-channel device, shown in Fig. 2 (a). We expect that for short-channel devices it is normal to observe the decrease of field-effect mobility and nonlinear scaling of drain current due to two reasons[9]. First is the contact resistance, which does not scale with channel length and starts to dominate at short channel lengths. Second is charge carrier velocity saturation, which occurs at smaller electric fields as the channel length is decreased, resulting in a smaller effective field effect mobility due to drain current saturation.

The cutoff frequency $f_T$, the frequency at which the current gain becomes unity is given by

$$f_T = \frac{g_m}{2\pi} \frac{1}{(C_{gs}+C_{gd})[1+g_{ds}(R_s+R_d)]+C_{gd}g_m(R_s+R_d)},$$

where $g_m$ is the intrinsic transconductance, $C_{gs}$ is the gate-source capacitance, $C_{gd}$ is the gate-drain capacitance, $g_{ds}$ the drain conductance, $R_s$ and $R_d$ the source and drain series resistances, respectively. In Fig. 2(c), we plot the current gain, calculated from S parameters; a peak cutoff frequency $f_T$ of 8 GHz after the de-embedding procedure is obtained for the 70-nm-long device at room temperature. We performed RF measurements at DC parameters $V_{ds}$ = 2.5 V and $V_{tg}$ = −700 mV, which gives us the highest values of intrinsic transconductance.

The available power gain is another important figure of merit for RF transistors. The frequency at which the power gain is equal to unity, is called the maximum frequency of oscillation $f_{max}$,

$$f_{MAX} = \frac{f_T}{2\sqrt{g_{ds}(R_g+R_s)+2\pi f_T C_g R_g}}$$

where $g_{ds}$ is the drain differential conductance, $R_s$ the source resistance, $C_{gd}$ the gate to drain capacitance and $R_g$ is the gate resistance which mainly depends on the gate thickness and area.

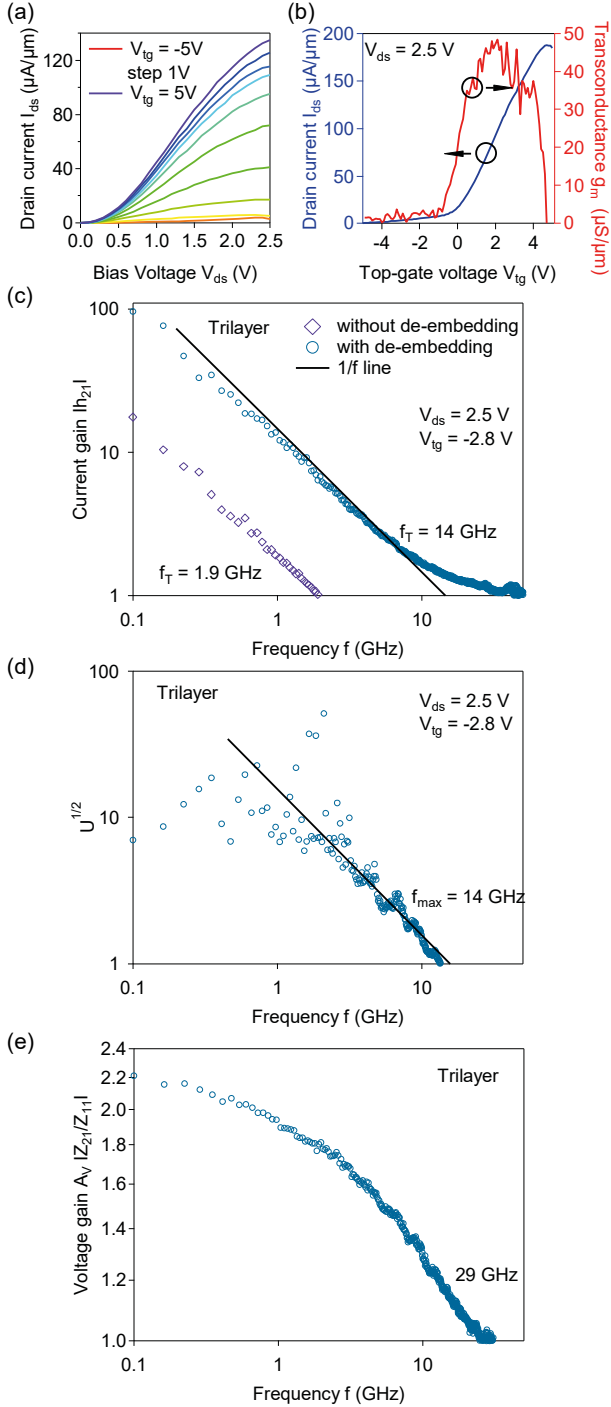

**Fig. 3** Device characteristics for 40 nm gate length and conventional source-drain contacts. (a) The drain current $I_{ds}$ as a function of bias voltage $V_{ds}$ for the trilayer MoS$_2$ device, measured for different top – gate voltages. (b) Left: Transfer characteristics for the trilayer MoS$_2$ FET under the applied drain frequency for the device with a 40 nm gate length based on trilayer MoS$_2$. The maximum frequency of oscillations $f_{max}$ = 12 GHz. (e) Voltage gain $|Z_{21}/Z_{11}|$ as a function of frequency for device with a 40 nm gate length based on trilayer MoS$_2$, showing gain up to 29 GHz.

In Fig. 2(d) we present Mason's unilateral power gain U as a function of frequency for the 70 nm gate length MoS$_2$ device with $f_{max}$ = 12 GHz. We also extract the intrinsic voltage gain $A_v = g_m/g_{ds}$ for the 70 nm gate length device by converting the scattering S-parameters to impedance Z-parameters where $A_v = Z_{21}/Z_{11}$. The result is presented in Fig. 2(e).

We have also analyzed the performance of 40 nm gate length trilayer MoS$_2$ devices. Fig. 3 (b) shows the transfer characteristics for 40 nm gate length MoS$_2$ with the intrinsic transconductance of 50 µS/µm. From the $I_{ds} - V_{tg}$ characteristic of our device, we extract estimates for the field effect mobility µ = 2.3 cm$^2$/Vs and contact resistance $R_c$ = 263 Ω or $R_c$ = 2.9 kΩ·µm with the width of the channel of 11.2 µm. Fig. 3 (a) presents the output behavior of trilayer MoS$_2$ device with a 40 nm gate length with the pronounced saturation of the drain-source current.

The cutoff frequency, obtained by extrapolation to a −20 dB/dec slope expected for conventional FETs is $f_T$ = 14 GHz, Fig. 3(c). Predictably, this value is higher compared to 70 nm gate length devices due to decreased electron transit time in shorter gate length devices. What was very interesting to find out is that with by decreasing the gate length down to 40 nm we were still able to improve the performance of the power gain with $f_{max}$ = 14 GHz. This is the missing point in RF graphene devices due to the absence of the band gap[8]. The voltage gain $A_v$ for 40 nm gate length is obtained until 29 GHz, Fig. 3(e).

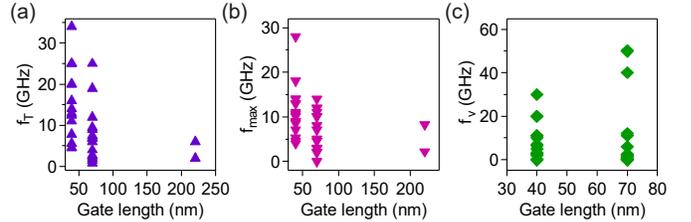

**Fig. 4** (a) Scaling behavior of $f_T$ as a function of channel length, showing 1/L dependence. (b) Scaling behavior of $f_{MAX}$ as a function of channel length, showing 1/L dependence. (c) Scaling behavior of the cutoff frequency for voltage amplification $f_V$ as a function of channel length.

While exhibiting considerable spread over various test devices, both $f_T$ and $f_{max}$ show an increasing tendency for smaller gate lengths, Fig. 4(a),(b). The voltage gain drops with decreasing the gate length, Fig. 4(c) which could be related to the carrier velocity saturation. In addition, the contact resistance becomes dominant compared to channel resistance in short channel devices[9], making further scaling difficult. To address this issue, we fabricated a batch of devices with cuts introduced in the MoS$_2$ film so that the MoS$_2$ edge length available for bonding with the contact metal is maximized, Fig. 1(b).

Trilayer MoS$_2$ FETs with cuts have µ = 6 cm$^2$/Vs and contact resistance of $R_c$ = 1.8 kΩ·µm or $R_c$ = 61 Ω with the 30 µm channel width, extracted from the transfer characteristics[1] presented in Fig. 5(b). The maximum intrinsic transconductance reaching the value of 100 µS/µm, right side of Fig. 5 (b). From Fig. 5 (a) we can see the saturation in the drain-source current, which results in voltage gain up to 45 GHz, Fig. 5(e). For this device type, we obtained the record value of $f_T$ = 6 GHz before the de-embedding procedure and 25 GHz after the de-embedding procedure, Fig. 5 (c) and the highest value of $f_{max}$ = 16 GHz.

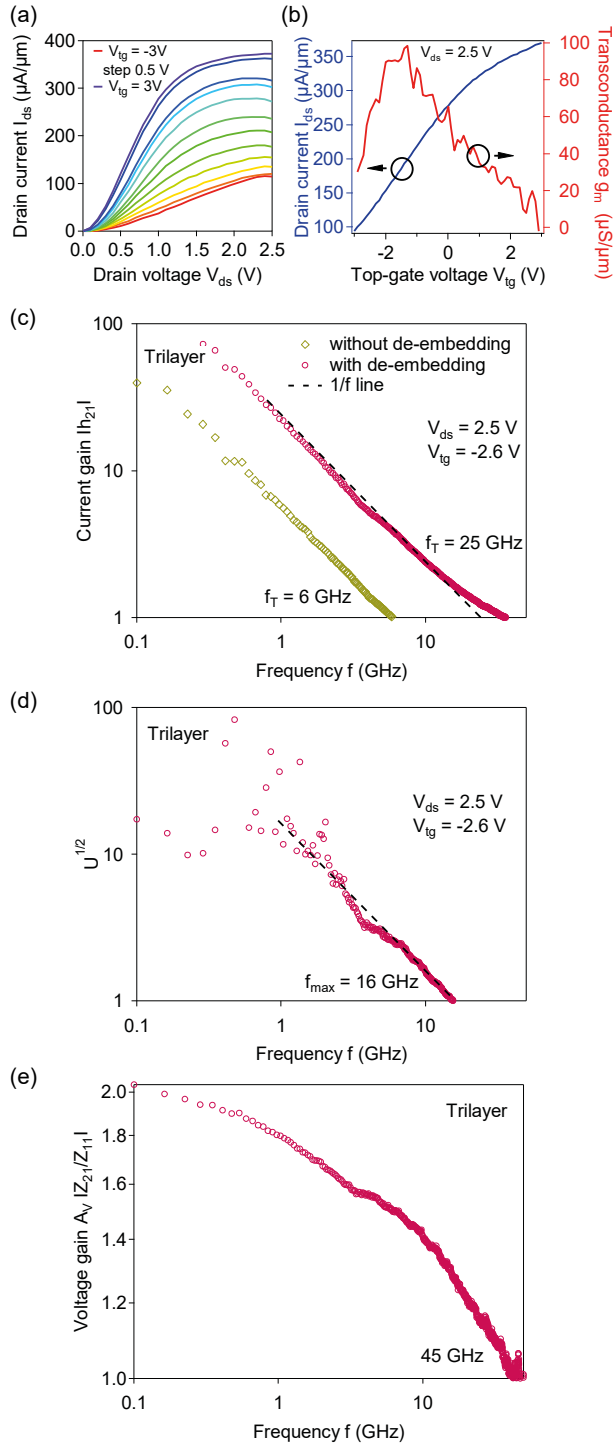

**Fig. 5** Device characteristics for 70 nm gate length and patterned contact area for reduced contact resistance. (a) $I_{ds} - V_{ds}$ characteristic for the trilayer $MoS_2$ device, measured for different top-gate voltages. (b) Left: transfer characteristics for the trilayer $MoS_2$ FET under the applied drain voltage $V_{ds} = 2.5$ V. Right: transconductance of trilayer $MoS_2$ (c) Small-signal current gain $h_{21}$ as a function of frequency for a device with 70 nm gate length based on trilayer $MoS_2$ with the patterned contact area. The cut-off frequency $f_T = 25$ GHz. (d) Mason's unilateral gain U as a function of frequency. The maximum frequency of oscillations $f_{max} = 16$ GHz. (e) Voltage gain $|Z_{21}/Z_{11}|$ as a function of frequency for the device with a 70 nm gate length based on trilayer $MoS_2$ with the patterned contact area, showing voltage gain up to 45 GHz.

## Conclusion

In this work we have performed the analysis of $MoS_2$ FETs with gate lengths of 70 nm and 40 nm. We reach a cutoff frequency for voltage gain of 50 GHz in the case of trilayer $MoS_2$ FET with a 70 nm gate length. An analysis of $MoS_2$ RF transistors in terms of current, power and voltage gains was carried out. In order to decrease the contact resistance and to improve the properties of our devices in RF range we introduced the "edge – contacted" injection in trilayer $MoS_2$ RF-FETs. With this approach we boost the performance of our devices to the values of $f_T$ = 25 GHz, $f_{max}$ = 16 GHz and voltage amplification up to 45 GHz for trilayer $MoS_2$.

## Acknowledgements

Device fabrication was carried out in the EPFL Center for Micro/Nanotechnology (CMI). We thank Z. Benes (CMI) for technical support with e-beam lithography. This work was financially supported by Swiss SNF Grants 135046 and 144985.